\documentclass[12pt]{article}
\usepackage{epsfig}
\def\be{\begin{eqnarray}}
\def\ee{\end{eqnarray}}
\def\lb{\label}

\renewcommand{\theequation}{\arabic{section}.\arabic{equation}}
\renewcommand{\thesection}{\arabic{section}.}

\catcode`@=11 \@addtoreset{equation}{section} \catcode`@=12
\mathchardef\SGamma="7100

\begin{document}
\vskip 0.5cm
\hfill {\tt hep-th/0307011}\\
\vskip 0.4cm
\begin{center}
\vskip 2truecm {\Large\bf Echoing  the extra dimension
}\\
\vskip 1truecm {\large\bf A. O. Barvinsky${}^1$\footnote{\tt
barvin@lpi.ru}  and Sergey N.~Solodukhin${}^2$\footnote{ {\tt
soloduk@theorie.physik.uni-muenchen.de}}
}\\
\vskip 0.6truecm
\it{$^1$
Theory Department, Lebedev Physics Institute, \\
Leninsky prospect 53,
Moscow 117924, Russia}
\vskip 0.4truecm
\it{${}^2$Theoretische Physik,
Ludwig-Maximilians Universit\"{a}t,\par
Theresienstrasse 37,
D-80333, M\"{u}nchen, Germany}
\vskip 1truemm
\end{center}
\vskip 1truecm
\begin{abstract}
\noindent

We study the propagating gravitational waves as a tool to probe
the extra dimensions. In the set-up with one compact extra
dimension and non-gravitational physics resigning on the
4-dimensional subspace (brane) of 5-dimensional spacetime we find
the Green's function describing the propagation of 5-dimensional
signal along the brane. The Green's function has a form of the sum
of contributions from large number of images due to the
compactness of the fifth dimension. Additionally, a peculiar
feature of the causal wave propagation in five dimensions  
(making a five-dimensional spacetime very much different
from the familiar four-dimensional case) 
is that the entire region inside
the past light-cone contributes to the signal at the observation
point. The 4-dimensional propagation law is nevertheless
reproduced at large (compared to the size of extra dimension)
intervals from the source as a superposition of signals from large
number of images. The fifth dimension however shows up in the form
of corrections to the purely 4-dimensional picture. We find three
interesting effects: a tail effect for a signal of finite
duration, screening at the forefront of
this signal and a frequency-dependent amplification for a periodic
signal. We discuss implications of these effects in the
gravitational wave astronomy and estimate the sensitivity of
gravitational antenna needed for detecting the extra dimension.

\end{abstract}
\newpage
\section{Introduction}
Over decades since the works of Kaluza and Klein the idea of extra
dimensions enjoys constantly growing interest of the physical
community. Designed originally to serve the unification of the
General Relativity and Maxwell electrodynamics it is now more
fashionable to invoke the extra dimensions for solving the
problems of hierarchy and vacuum energy. A recent review on old
and new paths to the extra dimensions is \cite{Rubakov}. It is
customary nowadays to consider the Kaluza-Klein idea in the
context of the brane paradigm. According to this picture, the
non-gravitational physics, conveniently described by the Standard
Model, is confined to a $(3+1)$-dimensional subspace called brane of
the $(4+n)$-dimensional space-time with $n$ compactified extra
dimensions. Only gravitational field is allowed to propagate
through the whole spacetime and probe the extra dimensions. If $L$
is a typical size of the extra dimensions the gravitational
physics on the brane is expected to be $(4+n)$-dimensional on the
scales of order $L$ and smaller while the familiar 4-dimensional
gravitational interaction is supposed to emerge on the much larger
than $L$ scales. A standard way to study this is to look at the
static Newton potential along the brane. At large distances $r\gg L$
it takes the 4-dimensional form $G_4/4\pi r$, with $G_4$ being the
induced 4-dimensional Newton constant, while the higher
dimensional physics manifests in exponentially small corrections,
$e^{-r/L}$.

In this paper we give up staticity and look at propagation of
waves. The higher-dimensional corrections to the purely
4-dimensional law of propagation then are no more exponentially
small but have a power law. This makes the waves a more convenient
tool for probing the extra dimensions. Some puzzling surprises are
however awaiting for us on this venue. Most surprising is the fact
that the causal propagation of wave signals from a source is
radically different for odd and even dimensions. This
difference\footnote{The mathematical theory of the propagation of
waves in various dimensions is known for quite long time, see for
example \cite{Shilov}.} is most transparent in the retarded
potentials. In even dimensions, $(3+1)$-dimensional space-time
being a particular example, the retarded potential is proportional
to the delta-function of $(t-r)$ and thus is concentrated on the
future light-cone and vanishing everywhere else.  In contrast, the
retarded potential in odd spacetime dimensions has support
everywhere inside the light-cone! Additionally, there appear
non-integrable divergences on the forefront of the propagating
wave. The above properties make the wave propagation in
5-dimensional space-time very much different from the propagation
in the 4-dimensional case. In particular, the night sky would  look
quite differently if we lived in 5-dimensional Universe. Even
already dead stars would still produce  flashing spots on the sky
so that night would not be as dark as in our home Universe. A
natural question arises\footnote{ This question should have been
asked years ago. However, with few exemptions
\cite{Galtsov:2001iv}, \cite{Frolov:2003mc}, \cite{Cardoso} this property of wave
propagation remains unknown in the current literature on extra
dimensions.}: how after all these differences the large scale
propagation on the brane still manages to be 4-dimensional even
though it is built out of the 5-dimensionally propagating waves.
In this paper we address this question and moreover study the
manifestations of the 5-dimensional character of the wave
propagation in the form of corrections to the purely 4-dimensional
law.

We find a few interesting effects. First, we consider propagation of
a signal of finite  duration $T$ emitted by a source on the brane.
In purely 4-dimensional picture an observer at some large distance
from the source would see a pulse of same finite duration $T$. In
the 5-dimensional picture with compact extra dimension the signal
detected by an observer is a superposition of signals coming from
large number of images additionally to the original source. In
fact, the purely 4-dimensional picture arises as a result of this
superposition. The corrections are however important. Most
significantly, when the entire signal has passed through the point
of observation there still remain signals coming from the images
which are inside the past light-cone of the point of observation.
These signals keep arriving and result in a tail, a power-like
decaying with time signal. The whole effects thus is due to the
combination of the 5-dimensional feature of the propagation inside
the light-cone which we mentioned above and the compactness of the
fifth dimension manifested in presence of large number of images.
On the other hand, the shape of the forefront of the signal 
is modified by what we call screening effect: corrections appear 
with negative sign and smooth out the otherwise a step-like forefront 
profile. Another effect appears for the periodic signal. We find that the
amplitude of the signal at the point of observation is amplified
with a frequency-dependent amplification factor. Both effects are
proportional to $L$ in some power and are clear indications of the
presence of the extra dimension. This makes it interesting to
reconcile these effects with modern observational possibilities
available at LIGO or LISA.

This paper is organized as follows. In section 2 we formulate the
set-up and find various representations for the Green's function
describing propagation of the 5-dimensional signal along the
lower-dimensional subspace (brane). We focus on the retarded
potentials in section 3 and derive the 5-dimensional potential
consistent with the boundary conditions we impose along the fifth
dimension. The propagation of the finite pulse is considered in
section 4 and the periodic signal in section 5. In section 6 we
estimate the predicted effects and compare them with the available
sensitivity of the gravitational detectors.

\section{Many faces of the Green's function}
\subsection{The set-up}
We consider flat 5-dimensional space-time covered by coordinates
$X^A=(x^\mu,y)$ where $y$ is the coordinate along the fifth
dimension and $x^\mu, ~\mu=0,1,2,3$ are the four-dimensional
coordinates. We restrict the coordinate $y$ to lie within the
interval between $y=0$ and $y=L$ and we impose the Neumann
boundary conditions on the perturbations of the five-dimensional
gravitational field at the ends of this interval. $L$ is thus a
size of the extra dimension. Our four-dimensional world with
experimental devices, standard model particles and human beings
resign on the four-dimensional subspace at $y=0$ so that only
gravitational field may propagate into the fifth dimension. The
subspace $y=0$ (as well as the one at $y=L$) can be also thought
of as a zero-tension brane. We sometimes will use this
terminology. A generalization to  branes with non-vanishing
tension (the Neumann boundary condition then has to be replaced by
a condition of the Robin type) is possible although technically
more involved. Another possible setting is to impose the periodic
boundary conditions $\phi(x,y=0)=\phi(x,y=L)$, where $\phi(x,y)$
stands for the relevant five-dimensional perturbation. In this
case  we have a standard Kaluza-Klein configuration. Although all
our results are naturally extended to the case with periodic
boundary conditions, we will stick to the scenario with the
Neumann boundary conditions.

\subsection{Kaluza-Klein representation}
It should be noted that in what follows we systematically ignore
the tensor structure of the gravitational perturbations assuming
that the field equation of interest takes the form of the scalar
type field equation
    \be
    \left(-\partial_0^2
    +\partial_y^2+\partial_{{\bf x}}^2 \right)
    \phi (y,x^0,{\bf x})=0~~ \lb{1.0}
    \ee
subject to the boundary conditions
    \be
    \partial_y \phi (y,x^0,{\bf x})|_{y=0}
    =\partial_y \phi
    (y,x^0,{\bf x})|_{y=L}=0~~. \lb{bc}
    \ee
The normalized modes forming the basis of functions in the
$y$-direction are
    \begin{eqnarray}
    &&\varphi_n(y)=\frac1{\sqrt{L}},\,\,\, n=0 \\
    &&\varphi_n(y)=\sqrt{\frac2L}\,\cos(m_n\,y),
    \,\,\,\,m_n=\frac{n\pi}L~~,\,\,\,n>0           \label{1.2}
    \end{eqnarray}
while in the $x$-direction the modes are the usual plane waves
$e^{ipx}$ appropriately normalized on the delta-function.
In terms of these modes the 5-dimensional propagator (or
the Green's function)
defined as a solution to the field equation with delta-like source
\be
-\nabla^2_5~D_5(X,X')=\delta(X,X')
\ee
takes the following five-dimensional momentum space representation
  \begin{eqnarray}
    D_5(X,X')=
    \frac1{(2\pi)^4}\int d^4p\,
    \sum\limits_{n=0}^{\infty}e^{ip(x-x')}\;
    \frac{\varphi_n(y)\,\varphi_n(y')}
    {p^2+m^2_n}   ~~,                        \label{1.1}
  \end{eqnarray}
where $p^2=-p_0^2+{\bf p}^2$. In Lorentzian signature there are
various types of propagator depending on the contour chosen when
the integration over momenta is taken. The particular choices of
the integration will be discussed in the next section. Here for
simplicity of the presentation  we switch to the Euclidean
signature by means of the analytic continuation $p_0=i p_4$ so
that $p^2$ becomes a positive quantity and the problem of the
integration contour does not arise.

We see that $m_n$ introduced in (\ref{1.2}) have the meaning of
the Kaluza-Klein masses. Note that in case of periodic boundary
conditions we would get a similar form for the propagator with
different set of Kaluza-Klein masses $m_n=2\pi n/L$.

The main object of interest in the present paper is the
reduction of the propagator (\ref{1.1}) to the subspace $y=0$.
In other words, we are interested in the situation when
the gravitational 5-dimensional signal propagates between two points
on the subspace where we happen to live. This propagation is described
by the quantity
  \begin{eqnarray}
    D(x,x')\equiv \left.D_5(X,X')\,
    \right|_{y=y'=0}               ~~.             \label{1.3a}
    \end{eqnarray}
It is clear from equation (\ref{1.1}) that $D(x,x')$
reduces to the series of 4-dimensional massive Green's functions of
the Kaluza-Klein tower
    \begin{eqnarray}
     D(x,x')=\frac1{(2\pi)^4 L}\int d^4p\,
    \sum\limits_{n=-\infty}^{\infty}
    \frac{e^{ip(x-x')}}
    {p^2+m^2_n}=\frac1L
    \sum\limits_{n=-\infty}^{\infty}
    D(m_n^2|\,x-x')~~,                       \label{1.3}
  \end{eqnarray}
where
    \be
    D(m^2|\,x-x')=\frac{m}{4\pi^2}
    \frac{K_1(m|x-x'|)}{|x-x'|},
    \,\,\,\,|x-x'|=\sqrt{(x-x')^2}     \lb{prop}
    \ee
is the 4-dimensional propagator of massive field given in
Euclidean space by the MacDonald function of the first order.
Equation (\ref{1.3}) gives us the first, very natural from the
Kaluza-Klein perspective, representation for the brane-to-brane
propagator as mediated by the set of massive Kaluza-Klein fields.

\subsection{Momentum space representation}
Another way to look at the expression (\ref{1.3}) is to consider
it as a momentum space representation for the entirely
four-dimensional propagator
   \be
   D(x,x')=\frac1{(2\pi)^4 L}\int d^4p\,
   D(p)\, e^{ip(x-x')}                         \lb{pp}
   \ee
with the non-trivial propagator $D(p)$ as a function of the
4-dimensional momentum $p$. $D(p)$ is expressed in terms of the
infinite sum over Kaluza-Klein masses as in (\ref{1.3}). For the
tower of these masses given by (\ref{1.2}) (tracing back to the
boundary conditions (\ref{bc})) the sum over $n$ can be calculated
explicitly and the answer is
  \begin{eqnarray}
     D(p)\equiv \sum\limits_{n=-\infty}^{\infty}
    \frac1{p^2+m^2_n}
    =\frac{L}p\frac{\cosh(Lp)}{\sinh(Lp)},
    \,\,\,p=\sqrt{p^2}       ~~.                  \label{1.4}
  \end{eqnarray}
Notice that (\ref{1.4}) is valid for the Euclidean signature.
Switching to the Lorentzian signature is provided by the analytic
continuation $p\rightarrow ip$ so that the Lorentzian form of the
propagator $D(p)$ at timelike momenta is the following
     \be
     D_M(p)={L\over p}{\cos (Lp)\over \sin(Lp)},
     \,\,\,p=\sqrt{-p^2}~~.                       \lb{dp}
     \ee
Notice that it has poles at real values of $p$, $L\,p=\pi\,n$,
which correspond to the Kaluza-Klein modes. Additionally $D_M(p)$
has zeros at $Lp={\pi\over 2}(2n+1)$, $n$ is integer. The physical
interpretation of the zeros of the propagator is less
clear\footnote{These zeros guarantee the positivity of residues of
the propagator at its poles or the normal non-ghost nature of the
corresponding Kaluza-Klein modes and also follow from the duality
of the Dirichlet and Neumann boundary value problems in braneworld
physics \cite{duality}.}. At large values of momentum (or,
respectively for very short space-time intervals) the propagator
(\ref{1.4}) behaves as $D(p)\sim{1\over p}$, the typical behavior
of truly 5-dimensional propagation as it is seen from a
lower-dimensional subspace. On the other hand, at small values of
momentum (i.e. at large intervals on the brane) the
four-dimensional behavior $D(p)\sim {1\over p^2}$ is recovered.
This guarantees us that the four-dimensional physics is reproduced
at large scales even though the underlying physics is
5-dimensional\footnote{For a curved de Sitter brane embedded in various
5-dimensional backgrounds this transition was analyzed in detail in 
\cite{plb}}. The cross-over between two regimes happens at the
space-time intervals $s^2(x,x')$ of order $L^2$. Provided we were
not aware that our four-dimensional world is embedded into a
higher-dimensional space-time we would have to deal with the
propagator (\ref{dp}) as a non-trivial function of the 4-momentum
$p$ not being aware about its Kaluza-Klein origin.

\subsection{Sum over images representation}
The integration over momentum in (\ref{pp}) includes the integration
over angles and the integration over absolute value $p$ of the 4-momentum.
Inserting the exact expression (\ref{1.4}) into (\ref{pp}) the
integration over angles can be performed explicitly and gives
  \begin{eqnarray}
    D(x)=
    \frac1{4\pi^2}\frac1{|x|}\int\limits_0^\infty
    dp\,p\,{1\over L}\, D(p)
\,J_1(p|x|)~~,  \label{1.5}
  \end{eqnarray}
where the table integral
  \begin{eqnarray}
    \int\limits_0^\pi d\theta\,\sin^2\theta\,
    e^{iy\cos\theta}=\frac\pi{y}\,J_1(y)
  \end{eqnarray}
has been used. Taking into account that on the real axis the
following relation holds between the Bessel and Hankel functions
  \begin{eqnarray}
    J_1(z)=\frac12\left(H^{(1)}_1(z)
      +H^{(2)}_1(z)\right)=
    \frac12\left(H^{(1)}_1(z)+H^{(1)}_1(-z)\right)  \label{1.6}
  \end{eqnarray}
one can transform the integral to the integration over the whole real
axis of $p$ with the principal value prescription at $p=0$
  \begin{eqnarray}
   \int\limits_0^\infty
    dp\,p\,
{1\over L}\,D(p)
\,J_1(p|x|)
    =\frac12\,{\rm P}\int\limits_{-\infty}^\infty
    dp\,p\,
{1\over L}\,D(p)
\,H^{(1)}_1(p|x|)~~.   \label{1.7}
  \end{eqnarray}
By closing the contour of integration in the upper-half plane of the complex
variable $p$ one finds that the result reduces to the contribution of poles
on the imaginary semiaxis, $p_n=im_n$, $n\geq 0$ (the pole at $n=0$
contributing only one half of the residue in view of the principal value
integration above)
  \begin{eqnarray}
    \int\limits_0^\infty
    dp\,p\,\frac{\cosh(Lp)}{\sinh(Lp)}
    \,J_1(p|x|)=\frac1{L|x|}
    +\frac{2\pi}{L^2}\,
    \sum\limits_{n=1}^\infty n\,K_1(m_n|x|)~~,     \label{1.8}
  \end{eqnarray}
where $K_1(z)=-\pi H^{(1)}_1(iz)$ is the modified Bessel (MacDonald) function.
In view of the relation $K_1(z)=-\partial_z K_0(z)$ between the MacDonald
functions of the zeroth and first order, one has
  \begin{eqnarray}
   \int\limits_0^\infty
    dp\,p\,\frac{\cosh(Lp)}{\sinh(Lp)}\,J_1(p|x|)=\frac1{L|x|}
    -\frac{2\pi}{L^2}\,\partial_z\!
    \sum\limits_{n=1}^\infty K_0(nz)\,
    \Big|_{z=\pi|x|/L}~~,                    \label{1.9}
  \end{eqnarray}
where the series of McDonald functions can be calculated with the aid of the
formula \cite{GradRyzh}
  \begin{eqnarray}
    \sum\limits_{n=1}^\infty K_0(nz)=\frac12
    \left({\mathbf{C}}+\ln\frac{z}{2\pi}\right)+\frac\pi{2z}+
    \pi\sum\limits_{n=1}^\infty
    \left(\frac1{\sqrt{z^2+4\pi^2n^2}}
      -\frac1{2\pi n}\right)~~.                \label{1.10}
  \end{eqnarray}
Putting things together this gives us yet another, coordinate space,
representation for the propagator
  \begin{eqnarray}
     D(x,x')=\frac1{4\pi^2}
   \sum\limits_{n=-\infty}^{\infty}
   \frac1{((x-x')^2+4n^2\,L^2)^{3/2}}~~.              \label{1.11}
  \end{eqnarray}
This result obviously reproduces the recovery of the  Green's
function along the brane
by another method -- method of images. Indeed, it can be obtained by
summing the contributions of images of the original source, located at the
infinite sequence of points $X_n$
  \begin{eqnarray}
    &&D(x)=2\sum\limits_{n=-\infty}^{\infty}
    D_5(X-X_n)\,\Big|_{\,X=(x,0)},  \\
    &&X_n=(0,\,2nL),\,\,\,\, (X-X_n)^2=x^2+4n^2L^2~~,  \label{1.12}
  \end{eqnarray}
mediated by the five-dimensional Green's function\footnote{The coefficient
of two here arises from the fact that the magnitude of the source effectively
doubles when the source approaches the brane and merges with one of its
primary images -- this happens with all pairs of higher-order images merging
at $X_n=(0,2Ln)$, $n\neq 0$.}
  \begin{eqnarray}
    D_5(X,X')
    =\frac1{8\pi^2}\frac1{((X-X')^2)^{3/2}}~~.             \label{1.13}
  \end{eqnarray}
The representation (\ref{1.11}) of the propagator expresses the
obvious fact that there many ways  to propagate signal between two
space-time points on the brane: it may propagate entirely along
the brane (the $n=0$ term in the sum (\ref{1.11})) or through the
fifth dimension reflecting off the second brane at $y=0$ and
coming back to our brane. Each such propagation contributes to the
sum in (\ref{1.11}). Since the number of reflections can be
arbitrary high the sum has infinite number of terms.

When $(x-x')^2$ is much larger than $L^2$ the sum in (\ref{1.12})
can be approximated by the integral, $\sum_n={1\over 2L}\int \,
dy$ where $y=2\,L\,n$. Evaluating the integral we find
    \be
    D(x,x')\simeq {1\over 8\pi^2 L}
    \int_{-\infty}^{+\infty}dy\,
    {1\over ((x-x')^2+y^2)^{3/2}}=
    {1\over 4\pi^2 L}\,{1\over
    (x-x')^2}, \lb{ds}
    \ee
so that 4-dimensional Green's function is recovered in this limit.
It is actually a realization of the well-known in mathematics
method called descent method to obtain Green's function from a
Green's function one dimension higher. Thus, the 4-dimensional
propagator is recovered for large (compared to $L$) intervals on
the brane as a result of superposition of contributions from
infinite number of images.

\subsection{Newton potential}
We finish this section with a brief illustration of how the Newton potential
emerges from our analysis. Indeed, the Newtonian potential on the brane
can  be easily recovered from (\ref{1.11}) by integrating over time
(it is again a realization of the descent method discussed
in the previous subsection)
  \begin{eqnarray}
    V(r)=
    \int\limits_{-\infty}^{\infty} dx^4\,
    D({\mathbf x},x^4)
    =\frac1{2\pi^2}\sum\limits_{n=-\infty}^{\infty}
    \frac1{r^2+4n^2L^2}~~,~~r^2={\bf x}^2~~.          \label{1.14}
  \end{eqnarray}
In view of the summation formula (\ref{1.4}) it equals
  \begin{eqnarray}
     V(r)
     =\frac1{4\pi r}
     \left(1+\frac2{e^{\pi r/L}-1}\right)          \label{1.15}
  \end{eqnarray}
and, thus, reproduces at large distances $r\gg L$ the usual four-dimensional
Newtonian
potential with Yukawa-type corrections in the four-dimensional world. For
small distances $r\ll L$ it obviously features the five-dimensional
behavior $\sim 1/r^2$.

\section{Retarded potentials}
The retarded Green's function which is an object of our prime interest here
can be obtained from the Euclidean Green's functions of the previous section
by the analytic continuation. This can be done directly in the final
result (\ref{1.11}) in the coordinate representation, or separately for
massive
Green's functions of the Kaluza-Klein tower in (\ref{1.3}). Here we choose the
first method, and its equivalence to the second one is shown in the Appendix A.

To distinguish the coordinates of the Lorentzian spacetime
$x=(x^0,{\mathbf x})$ from those of the Euclidean space we supply the latter
by the subscript $E$, $x_E=(x^4,{\mathbf x})$. They are related by the Wick
rotation
  \begin{eqnarray}
   &&x_E\equiv(x^4,{\mathbf x})=(ix^0,\,{\mathbf x}),  \label{2.1}\\
   &&x_E^2\equiv(x^4)^2+{\mathbf x}^2=
   -(x^0)^2+{\mathbf x}^2+i\varepsilon=x^2+i\varepsilon~~,  \label{2.1a}
  \end{eqnarray}
which in particular determines the analytic continuation from the (massive)
Euclidean Green's function
  \begin{eqnarray}
 (m^2-\nabla^2_E) D_E(x_E)=\delta(x_E)           \label{2.2}
  \end{eqnarray}
to the Feynman propagator in Lorentzian spacetime -- the boundary
value on the real axis of the Lorentzian spacetime interval $x^2$
when approaching from the upper half complex plane of $x^2$
according to (\ref{2.1a}), $\varepsilon\to 0$, \cite{DW}
  \begin{eqnarray}
   (m^2-i\varepsilon-\nabla^2) D_F(x)=\delta(x)~~,~~D_F(x)=
    i D_E(x_E)\,\Big|_{\,x_E=(ix^0,\,{\mathbf x})}    \label{2.3}
  \end{eqnarray}
(the factor $i$ here follows from the formal identification
$\delta(x)=i\delta(x_E)$ matching with (\ref{2.1})).

As is well known \cite{DW}, other Green's functions in flat spacetime
can be obtained from the Feynman propagator by taking its real and
imaginary parts and multiplying it with distributions
reflecting the needed causality properties. In particular, the retarded
propagator reads
  \begin{eqnarray}
    D_R(x)=2\,\theta(x^0)\,{\rm Re}\,D_F(x)   ~~.       \label{2.4}
  \end{eqnarray}
We illustrate this prescription for massless field propagator.
Starting
with the massless Euclidean Green's function in four-dimensions
  \begin{eqnarray}
     D_E(x_E)\Big|_{\,m^2=0}=\frac1{4\pi^2}\frac1{x_E^2}
  \end{eqnarray}
one finds in view of (\ref{2.2}) and (\ref{2.4}) the well-known
retarded Green's function of a massless field
  \begin{eqnarray}
    D_R(x)\Big|_{\,m^2=0}
    =\frac{\theta(x^0)}{2\pi^2}\,{\rm Re}\frac{i}{x^2+i\varepsilon}
    =\frac{\theta(t)}{4\pi}
    \frac{\delta(r-t)}r,
    \,\,\,r=|{\mathbf x}|~~,\,\,\,t=x^0~~,      \label{2.5}
  \end{eqnarray}
where in the last line the well-known formula
$$
{1\over x^2+i\varepsilon}-{1\over x^2-i\varepsilon}
=-2\pi i \delta(x^2)
$$
has been used.
Important feature of this propagator in four dimensions is that it has a
support on the past light cone of the observation point.
As we shall see in a moment the behavior of the retarded potential in five
dimensions is rather different.

Application of this rule to 5-dimensional Green's function (\ref{1.13})
or (\ref{1.11}) is trickier because of the branching
points in this expression. Wick rotation for the Euclidean Green's function
(\ref{1.11}) is based on the relation
  \begin{eqnarray}
    (x^2\pm i\varepsilon)^\lambda=
    |x^2|^\lambda\,
    e^{\pm i\pi\lambda\theta(-x^2)}~~,
    \,\,\,\,|\lambda|<1~~.                \label{2.6}
  \end{eqnarray}
Because of the restriction on $\lambda$ we cannot apply the above
formula immediately to (\ref{1.13}) in which $\lambda=3/2$.
Therefore,  we first  effectively lower the degree of the
singularity by noting that
  \begin{eqnarray}
   \frac1{(x^2+i\varepsilon)^{3/2}}=
     -\frac1r\,\frac\partial{\partial r}
     \frac1{(r^2-t^2+i\varepsilon)^{1/2}}~~   \label{2.7}
  \end{eqnarray}
and then apply (\ref{2.6}) to get
  \begin{eqnarray}
    D_R^{(5)}(r,t)={1\over 4\pi^2}   {\rm Re}\,\frac{i}{(r^2-t^2+i\varepsilon)^{3/2}}=-{1\over 4\pi^2 r}
     \,\frac\partial{\partial r}
     \frac{\theta(t-r)}
     {(t^2-r^2)^{1/2}}~~.   \label{2.8}
  \end{eqnarray}
We should apply now this formula to each term in (\ref{1.11})
so that finally the kernel of the retarded Green's function takes the form
 \begin{eqnarray}
    D_R(t,{\mathbf x})=
     -\frac1{2\pi^2}\!
     \sum\limits_{n=-\infty}^{\infty}\frac1r\,
     \frac\partial{\partial r}\,
     \frac{\theta(t-\sqrt{r^2+4n^2L^2})}
     {\sqrt{t^2-r^2-4n^2 L^2}}~~.            \label{2.9}
  \end{eqnarray}
An alternative derivation of this result, starting directly with
retarded propagators for massive Kaluza-Klein modes in
(\ref{1.3}), is given in the Appendix A. Two drastic differences
from the 4-dimensional case (\ref{2.5}) should be mentioned.
First, the retarded Green's function (\ref{2.8}), (\ref{2.9}) has
support on entire region inside the future light-cone (in
4-dimensional case the support was just on the surface of the
light-cone) of the delta-like source at $r=0,\ t=0$. 
Second, the presence of non-integrable singularity on
the light-cone in (\ref{2.8}) should be noted. Additionally, all
images of the original source, belonging to the interior of the
past light cone of the observation point, contribute to the
Green's function (\ref{2.9}).

\section{Finite pulse}
The brane-to--brane Green's function (\ref{2.9}) gives on the
brane $y=0$ the solution of the equation
  \begin{eqnarray}
    \nabla^2_5\Phi(X)=-G_5\,J(X)    \label{2.10}
  \end{eqnarray}
with zero initial conditions in the past and with the source $J(X)$ located
on the same brane
  \begin{eqnarray}
  J(x,y)=\,f(t)\,\delta^{(3)}({\mathbf x})\,\delta(y)~~. \label{2.11}
  \end{eqnarray}
Here $G_5$ denotes the five-dimensional gravitational coupling constant.
This solution reads
  \begin{eqnarray}
    \Phi(t,{\mathbf{x}})=G_5
    \int\limits_{-\infty}^\infty dt'\,
     D_R(t-t',{\mathbf x})\,f(t')~~.   \label{2.12}
  \end{eqnarray}
Substituting (\ref{2.9}) we have
  \begin{eqnarray}
    \Phi(t,{\mathbf{x}})=-\frac{G_5}{2\pi^2}
    \sum\limits_{n=-\infty}^{\infty}
    \frac1{r_n}\,\frac\partial{\partial r_n}
    \int\limits_{-\infty}^{t-r_n} dt'\,\left.
    \frac{f(t')}{\sqrt{(t-t')^2-r_n^2}}\,
    \right|_{\,r_n=(r^2+4n^2L^2)^{1/2}}~~.    \label{2.13}
  \end{eqnarray}

When the source $f(t)$ in (\ref{2.11}) has the form of a constant pulse of
a finite duration $T$,
 \begin{eqnarray}
  f(t)=\theta(t)-\theta(t-T)~~,    \label{2.14}
  \end{eqnarray}
the expression (\ref{2.13}) (after the integration over $t'$ has been performed)
takes the form
  \begin{eqnarray}
    \Phi(t,{\mathbf{x}})=
    \frac{G_5}{16\pi^2L^3}\,t\,\theta(t-r)\,
    I\left(\alpha,\beta\,|\,
      [\alpha]\right)-(t\to t-T)~~,\label{2.15}
    \end{eqnarray}
in terms of the function $I(\alpha,\beta\,|\,[\alpha])$ of the two parameters
     \begin{eqnarray}
    &&\alpha=\frac{\sqrt{t^2-r^2}}{2L}~~,   \label{2.15a}\\
    &&\beta=\frac{r}{2L}~~.                  \label{2.15b}
  \end{eqnarray}
Here the square brackets in the notation $[\alpha]$ denote the
integer part of $\alpha>0$ and the
function of three arguments $I(\alpha,\beta\,|\,m)$ is given
by the following finite sum
  \begin{eqnarray}
    I(\alpha,\beta\,|\,m)=\sum\limits_{n=-m}^{m}
    \frac1{(\beta^2+n^2)\sqrt{\alpha^2-n^2}}~~,
    \,\,\,\,m<\alpha~~.                         \label{2.16}
  \end{eqnarray}

Let us first discuss the contribution of the forefront of the pulse --
the first term of (\ref{2.14}). With $t$ reaching the value of the
radial coordinate $r$ at which the observer is located, the observer starts
getting the signal encoded in the first term of (\ref{2.15}). As long as
this term depends on the integer part of the parameter $\alpha$, $m=[\alpha]$,
this signal is not continuous in time. For any value of time when $\alpha$
becomes integer, a new pair of terms at $n=\pm[\alpha]$ appear in the sum
(\ref{2.16}) and the value of the sum changes discontinuously. This
corresponds to the fact that the signals from the two new images located at
$y=\pm 2[\alpha]L$ reach the observer. Moreover, immediately at
$\alpha=[\alpha]$ these signals are singular, because they are proportional
to $1/\sqrt{\alpha-[\alpha]}$. This is a direct consequence of the structure
of singularity of the retarded Green's function in five dimensions. In contrast
to four dimensions its support is not restricted to the past light cone,
rather it is given by the entire interior of the latter (see Eq.(\ref{2.8})),
and the Green's function kernel in (\ref{2.8}) has a singularity
$\sim 1/(t^2-r^2)^{3/2}$ at $t\to r+0$. Because of the square root the
corresponding singularity in the $m=\pm[\alpha]$ terms of (\ref{2.16}) is
not strong, and when averaged over the time lapse between two consequitive
singularities gives, as we show below, a finite contribution.

Immediately after $t$ reaches the value of $r$ only the zeroth
term of the sum (\ref{2.16}),
$1/\alpha\beta^2=8L^3/r^2\sqrt{t^2-r^2}$ contributes to the signal
(\ref{2.15}). This corresponds to the contribution of only the
source located at $y=0$ (and its image merging with the source
and, thus, doubling its value -- see discussion in Sect.2).
Therefore this contribution is essentially 5-dimensional by its
nature and reflects the fact that the retarded potential
(\ref{2.8}) in five dimensions has non-integrable singularity on
the lightcone. As time goes on more signals from images start to
arrive so that at sufficiently large time the signal detected on
the brane has the shape of a sequence of singular spikes.
Obviously, an observer on the brane with devices at hand is not
capable to resolve the spikes so that the time averaging which we
employ below in the paper becomes a meaningful procedure.

The recovery of the 4-dimensional nature of the signal for a given
$r$ takes place after some time when the signals from many images
reach the observer and form the cumulative effect of many terms in
the sum (\ref{2.16}). This situation corresponds to the limit of
$\alpha\gg1$. The calculation of the sum (\ref{2.16}) in this
limit is presented in the Appendix B and gives the following
result in the first two orders of the asymptotic
$1/\alpha$-expansion
   \begin{eqnarray}
     &&I(\alpha,\beta\,|\,[\alpha])
     =\frac\pi\beta\,
    \frac{\coth\pi\beta}{\sqrt{\alpha^2+\beta^2}}
    -\frac{2^{3/2}}{\alpha^2+\beta^2}\,
    \sqrt{\frac{\delta}\alpha}\,
    \left[\,1-\frac1{\sqrt2}\,
      I(\delta)\,\right]_{\,\delta=\alpha-[\alpha]+1}\nonumber\\
    &&\qquad\qquad
    +\frac{2^{1/2}}{\alpha^2+\beta^2}\,
    \frac1{\sqrt{\alpha(\alpha-[\alpha])}}+
    {\rm O}\left(\frac1{\alpha^{7/2}}
     \right)~~.                               \label{2.24}
  \end{eqnarray}
Here $I(\delta)$ is the following integral of the parameter
$\delta=\alpha-[\alpha]+1$ belonging the finite range $1\leq\delta<2$
  \begin{eqnarray}
  I(\delta)=\int\limits_0^\infty
      \frac{dy}{e^{2\pi y\delta}-1}\,
      \left(\frac{(1+y^2)^{1/2}-1}
     {1+y^2}\right)^{1/2}.                  \label{2.22}
  \end{eqnarray}
This integral is a decreasing function bounded in this range by
its very small value at $\delta=1$, $I(1)\simeq 0.03$ and in what
follows gives a negligible contribution.

Similarly to the situation with small $\alpha$, the last term of
(\ref{2.24}) becomes singular also at large integer values of
$\alpha$, which corresponds to the arrival of the signal from the
new distant pair of images located at $|y|=2[\alpha]L\sim t\gg L$.
Interestingly, though, that the singular spikes occurring for late
times with the period $\Delta t\simeq4(\alpha
L^2/t)\Delta\alpha=2L\sqrt{1-r^2/t^2}\to 2L$ have now integrable
singularities, so that the averaging over time of this spiky
signal gives a finite average value. This averaging obviously
implies that the temporal resolution threshold of the observing
device is much bigger than the period $2L$ of spikes  for
hypothetic values of $L$ belonging to a millimetre range.

Thus, the temporal averaging of the spike singularity in (\ref{2.24})
is finite and equals
   \begin{eqnarray}
     \Big<\Big(\alpha-[\alpha]\Big)^{-1/2}\Big>
     \equiv\int\limits_{[\alpha]}^{[\alpha]+1}
     \frac{d\alpha}{\sqrt{\alpha-[\alpha]}}=2   \label{2.25}
  \end{eqnarray}
(we are averaging here only the most rapidly varying singular factor
while the other factors for
$\alpha\gg1$ can be regarded nearly constant). 
Similarly in the second term
of (\ref{2.24}) the averaging 
gives
   \begin{eqnarray}
     \Big<\sqrt{\alpha-[\alpha]+1}\Big>
     \equiv\int\limits_{[\alpha]}^{[\alpha]+1}
     d\alpha\,\sqrt{\alpha-[\alpha]+1}
     =\frac23\,\Big(2^{3/2}-1\Big)~~.            \label{2.26}
      \end{eqnarray}
Assembling these results together and taking into account the smallness
of $I(\delta)<0.03$ in (\ref{2.24}), one has the dominant
contribution to the temporal average of $I(\alpha,\beta\,|\,[\alpha])$
 \begin{eqnarray}
     \Big<I(\alpha,\beta\,|\,[\alpha])\Big>
     =\frac\pi\beta\,
    \frac{\coth\pi\beta}{\sqrt{\alpha^2+\beta^2}}
    +\frac{2^{3/2}}{\sqrt{\alpha}(\alpha^2+\beta^2)}\,
    \frac{5-2^{5/2}}3~~,\,\,\,\,\,\,\,\alpha\gg1     \label{2.27}
  \end{eqnarray}

Substituting this expression to (\ref{2.15}) and taking into account the
values of parameters (\ref{2.15a})-(\ref{2.15b}) we finally have the
retarded potential averaged over the time lapse between singular spikes
  \begin{eqnarray}
    &&\Big<\Phi(t,{\mathbf{x}})\Big>=
    \frac{G_4}{4\pi r}\,
    \left(1+\frac2{e^{\pi r/L}-1}\right)
    \,\theta(t-r)\nonumber\\
    &&\qquad-\frac{G_4}{3\pi^2t}\,
    \left(\frac{L^2}{t^2-r^2}\right)^{1/4}(2^{5/2}-5)\,
    \theta(t-r)\nonumber\\
    &&\qquad-(t\to t-T)~~,\,\,\,\,\,\,\,\,\,\,t^2-r^2\gg L^2~~. \lb{result}
  \end{eqnarray}
Here $G_4$ is an effective 4-dimensional gravitational constant
     \begin{eqnarray}
       G_4=\frac{G_5}L
     \end{eqnarray}
and the subtracted term at $t-T$ gives the contribution of the
back front of the pulse (\ref{2.14}) of finite duration $T$. It
should be noted that the first term in (\ref{result}) is exactly
the signal expected in 4-dimensions. Remarkably, the radial
function there is exactly the Newton potential which we calculated
in section 2. The second term in (\ref{result}) is a correction
which originated entirely due to the contribution of signals
arriving at a given point from large number of images.

When $t>T+r$ both signals from the forefront and the backfront (as
well as from their images) of the pulse start to arrive. As a
result, the first ``static'', i.e. dependent only on the radial
coordinate $r$ from the source, term cancels out. If we were in
4-dimensions this would be the only possible signal at $t>T+r$:
after the entire pulse has passed through our point our devices
detect no signal at all. In five dimensions the life is more
interesting: the whole region inside the lightcone contributes to
the signal detected at a given point.  So that even after the
shining candle dies off  one can still see its glimmering light.
In the present case an observer on the brane will detect a signal
coming not only from the source but also from a large number of
its images. As a result, we observe an interesting phenomenon:
even when the pulse has passed through the point of observation
our devices still detect some signal. This ``tail'' signal
originates  from the second term in  (\ref{result}) and takes the form
\be
\Big<\Phi(t,{\mathbf{x}})\Big>_{\rm tail} =(2^{5/2}-5){G_4\over 3\pi^2}
\left( {L^{1/2}\over (t-T)((t-T)^2-r^2)^{1/4}}-
{L^{1/2}\over t(t^2-r^2)^{1/4}}\right)~~.
\lb{tail0}
\ee
For later time when
$t\gg r$ and $t\gg T$ we have a simple expression for the tail
    \be
    \Big<\Phi(t,{\mathbf{x}})\Big>_{\rm tail}&=&
    (2^{5/2}-5)\, \frac{G_4}{2\pi^2 }\,
    \,{L^{1/2}\, T\over t^{5/2}}~~.             \lb{tail}
    \ee
It is proportional to the square root of the size of extra
dimension so that it is  a rather small quantity. It is also a
power-like decaying function of time and for late times, satisfying
the relation $t\gg r$, is distance-independent.
\begin{figure}
\centerline{\epsfig{figure=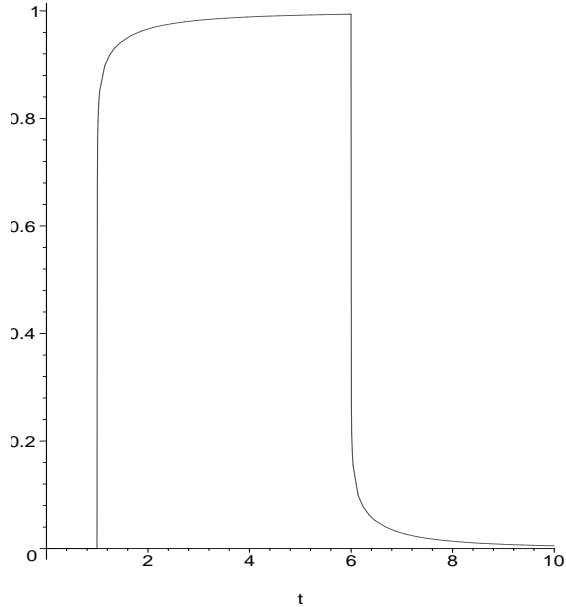,height=8cm}}
\caption{Typical shape of the signal (for illustration 
we take the initial profile to have unit amplitude
and duration $T=5$, ``size'' of the extra dimensions is $L=0.1$): forefront screening and 
the tail after backfront.}
\end{figure}

Finally, there is one more interesting 
effect: deformation of the forefront of the signal. It is due to the second
term in equation (\ref{result}). Noting the negative sign of this term
it effectively leads to  a relative screening  of the purely 4-dimensional
part of the signal. The typical shape of the resultant signal is shown in Fig.1.
In order to analyse this effect it is useful to rewrite eq.(\ref{result})
once again 
\begin{eqnarray}
    &&\Big<\Phi(t,{\mathbf{x}})\Big>\simeq
    \Phi_0(t,{\mathbf{x}})
    +\Phi_1(t,{\mathbf{x}}),      \label{00}\\
    &&\Phi_0(t,{\mathbf{x}})=
    \frac{G_4}{4\pi r},            \label{0}\\
    &&\Phi_1(t,{\mathbf{x}})=-\frac{G_4}{6\pi^2
    r}\,\frac{2^{5/2}-5}{\alpha (t)^{1/2}}=-\frac2{3\pi}
    \,\frac{2^{5/2}-5}{\alpha (t)^{1/2}}\;
    \Phi_0(t,{\mathbf{x}})~~,\,\,\,
\label{corr}
    \end{eqnarray}
where we assumed that  $r<t<r+T$ so
that no contribution of the backfront is yet present,
in terms of parameter\footnote{We neglect the 
difference between $\alpha$ and $[\alpha ]$
for large $\alpha$.} $\alpha(t)=(t^2-r^2)^{1/2}/2L$
having the meaning of number of spikes that have already passed 
through the observer by the time of observation.
This number is supposed to be sufficiently large to guarantee the 
formation of the four-dimensional signal (the first term in eq.(\ref{result})).
It also should be relatively small for the relation $t\sim r$
to hold. Eventually, the minimal value of $\alpha (t)$
is related to the time resolution of the measuring devise:
it may not resolve a single spike, but should resolve
the lapse between the arrival of the forefront and the
observation moment $t$.
Similar regime exists at the backfront tail: for $t$ close  to
$r+T$ the tail signal (\ref{tail0}) is 
dominated by the number of spikes arriving right after the original signal ends.
Later in the
paper we will discuss possible observational implications 
in the gravitational astronomy of the
tail effect (\ref{tail}) and, especially, the correction effect
(\ref{corr}).

\section{Periodic signal}
In this section we consider the situation when the source on the
brane radiates a periodic in time signal. So that we have
    \begin{equation}
    f(t)=e^{i\omega t}~~, \lb{5.1}
    \end{equation}
where $\omega$ is the frequency of the signal, to be substituted
in (\ref{2.13}). The $t$-integral then is taken explicitly to
produce
    \be
    \int_{-\infty}^{t-r}{dt'\, e^{i\omega t'}\over
    \sqrt{(t-t')^2-r^2}}= {(-\pi i)
    \over 2}\,e^{i\omega t}\,
    H^{(2)}_0(\omega r)~~,     \lb{5.2}
    \ee
so that the gravitational field takes the form
    \be \Phi(r,t)=
    -{G_5\over 8\pi i}e^{i\omega t}\,{1\over r}\,
    \partial_r \sum_{n=-\infty}^{+\infty}
    H^{(2)}_0(\omega\sqrt{r^2+4n^2L^2})~~. \lb{5.3}
    \ee
A possible way to estimate the infinite sum is to replace it by
integral. This approximation is valid when $\omega\, L<<1$. Then
we have that
    \be \sum_{n=-\infty}^{+\infty}
    H^{(2)}_0(\omega\sqrt{r^2+4n^2L^2})
    \simeq 2\int_0^\infty dx
    H^{(2)}_0(\omega\sqrt{r^2+4L^2x^2})
    ={1\over L\omega}\,e^{-i\omega
    r}~~,                              \lb{5.4}
    \ee
where in the passing to the last expression a table integral
available in \cite{GradRyzh} has been used. Thus, to the leading
order the gravitational signal is
    \be \Phi(r,t)\simeq
    \Phi_0(r,t)={G_4\over 8\pi r}\, e^{i\omega (t-r)}~~. \lb{5.4'}
    \ee
It is exactly the shape of the purely 4-dimensional signal arising
from a periodic source. Thus, by replacing the sum over images in
(\ref{5.3}) with the integral we have explicitly demonstrated our
earlier statement that the 4-dimensional physics is recovered from
5-dimensional as a superposition  of signals from infinite number
of images. In order to get a correction to the four-dimensional
result (\ref{5.4'}) we have to estimate a deviation of the
integral in (\ref{5.4}) from the sum. First, we estimate this
deviation for a single term in the sum (by expanding the integrand
to first term of its Taylor series),
    \be
    &&H^{(2)}_0(\omega\sqrt{r^2+4n^2L^2})-\int_n^{n+1}dx\,
    H^{(2)}_0(\omega\sqrt{r^2+4n^2x^2})\nonumber \\
    &&
=\int_n^{n+1}dx\,
    (H^{(2)}_0(\omega\sqrt{r^2+4n^2L^2})-
    H^{(2)}_0(\omega\sqrt{r^2+4L^2x^2}))\nonumber \\
    &&
\simeq{2\,n\,L^2\omega \over
    \sqrt{r^2+4L^2n^2}}H_1^{(2)}
    (\omega\sqrt{r^2+4L^2n^2})~~.        \lb{5.5}
    \ee
Summing now this deviation for all $n$ and again replacing the sum
by the integral (which turns out to be exactly integrable in view
of the relation $H_1^{(2)}(z)=(d/dz)H_0^{(2)}(z)$) we find
    \be
    \int_0^\infty dx {2xL^2\omega
    \over \sqrt{r^2+4L^2x^2}}
    H_1^{(2)}(\omega\sqrt{r^2+4L^2x^2})=
    {1\over 2} H^{(2)}_0(\omega r)~~.         \lb{5.6}
    \ee
Thus we find that the correction $\Phi_1(r,t)$ to the purely
4-dimensional result (\ref{5.4}) is
    \be
    \Phi_1(r,t)={G_4 \over
    8\pi i}{L\omega\over r}\,
    e^{i\omega t}\,H^{(2)}_1(\omega r)~~,    \lb{5.7}
    \ee
We are interested in a limit when the value of $\omega r$ is large.
Then we can use the asymptotic formula for the Hankel function and  get
    \be
    \Phi_1(r,t)\simeq {G_4 \over 8\pi }\,
    \sqrt{2\over \pi}\,
    {L\sqrt{\omega}\over   r^{3/2}}\,
    e^{i\omega (t-r)+i{\pi\over 4}}
    = \sqrt{\frac{2L^2\omega}{\pi r}}\,e^{i\pi/4}\,
    \Phi_0(r,t) ~~.    \lb{5.8}
    \ee
This part of the signal is entirely due to the extra dimension.
Notice the phase shift with respect to the 4-dimensional signal
(\ref{5.4'}). Also, the amplitude of (\ref{5.8}) now depends on
the frequency of the signal. These two circumstances help to
distinguish the purely 4-dimensional part of the signal from that
of due to presence of the extra dimension. 
We see that one of the manifestations of the fifth dimension is
some frequency-dependent amplification of the amplitude
$|\Phi_0(r,t)+\Phi_1(r,t)|$ of the signal from a periodic source.

\section{Can gravitational antenna detect the extra \\ dimension?}

Our analysis shows that there are two, at least in principle
observable, effects of the compact fifth dimension on the
propagation of the gravitational signal. First, it is the tail
effect discussed in section 4: after a signal of finite duration
$T$ has passed through the observer there still remains a tail
signal (\ref{tail}) entirely due to the presence of extra
dimension. The late-time tail signal is a power decaying function of time,
$t^{-5/2}$. In the four-dimensional physics the tail does not
appear at all. Therefore, its observation would be a clear
evidence for the fifth dimension. The amplitude of the purely
four-dimensional signal at distance $r$ from the source is the
usual Newton potential (\ref{0}), $\Phi_0(t,r)=G_4/4\pi r $.
Measuring the amplitude of the tail (\ref{tail}) with respect to
$\Phi_0$ we find that the upper bound on this ratio
is given by the quantity
    \be
    \chi\sim {L^{1/2}T\over r^{3/2}}\,
    ~~,  \lb{chi}
    \ee
where we took the observation moment $t\geq r$.

Another effect comes for a periodic signal of frequency $\omega$
from a source placed at distance $r$ from the point of
observation. The signal is amplified   due
to multiple winding around the compact extra dimension, moreover
the correction component (\ref{5.8}) has a phase shift useful for
observational purposes. The amplification factor is $1+\eta$ where
$\eta$ according to (\ref{5.8}) behaves as
    \be
    \eta\sim {L\over\sqrt{\lambda r}}~~, \lb{eta}
    \ee
in terms of the wavelength $\lambda=2\pi/\omega$ of the signal (in
our units the speed of light $c=1$).

The two effects suggest that the extra dimension in principle
could be observed in gravitational wave experiments and thus may
have some consequences for the gravitational astronomy. The
effects governed by parameters $\chi$ or $\eta$ are however
expected to be extremely small. Here we give some estimates
adjusted for the parameters (see for example \cite{LIGO} and 
\cite{Hughes:2002yy}) of the LIGO and LISA detectors. The
sensitivity of the LISA detector covers a low frequency range
$10^{-4}-1$ Hz. For a source placed at distance of 10 Mpc from the
Earth and producing a signal of finite duration $T\sim 10^4$ s
(associated with the inverse of the lower bound of this frequency
band) we have, assuming that the size of the extra dimension
$L\sim 10^{-1} $ cm, that $\chi\sim 10^{-25}$. The LIGO detector is
sensitive for a higher frequency, $1-10^4$ Hz, and is more
appropriate for the observation of the amplification effect for a
periodic signal. For the frequency $\omega\sim 100$ Hz and the
distance of 10 Mpc we find that $\eta\sim 10^{-18}$. 
The observation of the amplification effects thus requires less
amplitude sensitivity of the gravitational antenna.
We should not forget, however, that $\eta$ (or $\chi$) 
should be further multiplied by
the amplitude of the purely four-dimensional signal 
to give the value for the amplitude of the interesting part of the signal.
The resultant effect then is too small to be 
experimentally observable now or in coming future.
It is nevertheless worth noting that the late time tail (\ref{tail})
does not depend on the distance from the source. Therefore, there might be 
a cumulative effect of superposition of the tails from 
large number of various sources. The resultant signal is proportional to the
number of sources (both recent and in very distant past) 
in our part of the Universe and thus may have a
much larger amplitude and can be singled out by 
the characteristic $t^{-5/2}$ behavior.

There is yet another range of observation intervals which
gives more hope for experimental detection. This range corresponds
to (\ref{00})-(\ref{corr}) -- the transition regime from the
moment of the arriving forefront of the signal at $t=r$ to late
times $t> r$. In this regime the time is sufficiently larger
than $r$ to generate a large number $\alpha (t)$ of signal spikes  --
signals from images that already reached the observer -- but not
big enough to relax to negligible small tail rapidly decaying with
time. This number $\alpha(t)$ should be high enough to guarantee
the formation of the four-dimensional signal (\ref{00})
(remember that it gets completely recovered at infinite $t$ as a
cumulative effect of an infinite number of images), but it should
not be too big so that antenna would be able  to resolve 
the lapse after arriving the forefront and the observation time.
The correction to the 4-dimensional signal is given  by the product
$\alpha(t)^{-1/2}\Phi_0(t,{\mathbf{x}})$.
The lower bound on $\alpha$ is related to the time resolution
$\Delta t$ of the
antenna as $[{\rm min}~ \alpha ]\sim ({\Delta t\over L})^{1/2}~({r\over L})^{1/2}$.
This gives an estimate  $\alpha^{-1/2}\sim 10^{-8}$
(adjusted to the frequency $10^4$ Hz available at LIGO) 
and it makes certainly a much stronger effect than the ones  that have been just 
discussed.

Thus, the idea of echoing extra dimension(s) by the gravitational-wave
astronomy looks feasible, and we are planning to return to this
issue in subsequent publications.

\section*{Acknowledgements}
The authors are grateful to Slava Mukhanov for his stimulating
criticism. S.S. would also like to acknowledge a lively
discussion with Dejan Stojkovic and Andrei and Valeri Frolov.
A.O.B. is grateful for hospitality of the Physics Department of
LMU, Munich, where a major part of this work has been done under
the support of the grant SFB375. The work of A.O.B. was also
supported by the Russian Foundation for Basic Research under the
grant No 02-02-17054 and the grant LSS-1578.2003.2. Research of
S.S. is supported by the grant DFG-SPP 1096.

\newpage
\setcounter{section}{0}
\renewcommand{\theequation}{\Alph{section}.\arabic{equation}}
\renewcommand{\thesection}{Appendix \Alph{section}.}

\section{Alternative derivation of the retarded Green's function}
The retarded Green's function of the massive Klein-Gordon operator in four
dimensions
has a representation in terms of the Bessel function of zeroth order
\cite{retard,Rub}
\begin{eqnarray}
    G_R(m^2|\,t,{\mathbf x})=
     -\frac1{4\pi r}\,
     \frac\partial{\partial r}\,\Big[\,J_0(m\sqrt{t^2-r^2})
     \theta(t-r)\,\Big]~~.
  \end{eqnarray}
Therefore, the retarded version of the equation \ref{1.3} involves summation
over the tower of Kaluza-Klein masses $m_n=n\pi/L$. This summation can be
done with the aid of the Eq.(8.522) of \cite{GradRyzh}
  \begin{eqnarray}
     \sum\limits_{n=-\infty}^{\infty}J_0(n\pi s/L)
     =\frac{2L}\pi
     \sum\limits_{n=-m}^{m}\frac1{\sqrt{s^2-4n^2L^2}}~~,
  \end{eqnarray}
where $m$ is the integer satisfying the inequalities
  \begin{eqnarray}
    2 m<s/L<2(m+1)~~.
  \end{eqnarray}
This can be rewritten for $s=\sqrt{t^2-r^2}$ as
  \begin{eqnarray}
     \sum\limits_{n=-\infty}^{\infty}J_0(n\pi\sqrt{t^2-r^2}/L)
     =\frac{2L}\pi
     \sum\limits_{n=-\infty}^{\infty}
     \frac{\theta(t-\sqrt{r^2+4n^2L^2})}{\sqrt{t^2-r^2-4n^2 L^2}}~~,
  \end{eqnarray}
so that the resultant expression for the retarded Green's function
is given by equation (\ref{2.9}) in the main text.

\section{Calculation of $I(\alpha,\beta\,|\,m)$}
The calculation of the finite sum (\ref{2.16}) is based on the following
representation of this sum in terms of the contour integral in the
complex plane of the summation parameter $z$
  \begin{eqnarray}
    I(\alpha,\beta\,|\,m)=\frac\pi\beta\,
    \frac{\coth\pi\beta}{\sqrt{\alpha^2+\beta^2}}
      +\frac1{2i}\int\limits_{{\mathbf C}_m} dz\,
      \frac{\cot\pi z}
      {(\beta^2+z^2)\sqrt{\alpha^2-z^2}}~~.  \label{B.1}
  \end{eqnarray}
The closed contour of integration ${\mathbf C}_m$ here consists of the four
lines,
    \begin{eqnarray}
      &&{\mathbf C}_m={\mathbf C}_m^+\cup{\mathbf C}_m^-
      \cup{\mathbf C}^{+\infty}
      \cup{\mathbf C}^{-\infty},\nonumber\\
      &&{\mathbf C}_m^\pm:\,\,z=\pm m+iy,\,\,
      -\infty<y<\infty,\nonumber\\
      &&{\mathbf C}^{\pm\infty}:\,\,z=x\pm i\infty,\,\,
      -m<x<m~~,
      \end{eqnarray}
forming a rectangle gone around in anti-clockwise way. The contour integral
in (\ref{B.1}) gives the sum of contributions of residues at the poles of
$\cot\pi z$ which reproduce the needed finite sum (\ref{2.16}) and the sum of
two residues of the integrand at $z=\pm i\beta$ compensated by the first
term on the right hand side of (\ref{B.1}).

The integrals over ${\mathbf C}^{\pm\infty}$ obviously give zero
contribution, while the integrals over ${\mathbf C}_m^+\cup{\mathbf C}_m^-$
in view of the relation
   \begin{eqnarray}
     \cot\pi(\pm m+iy)=-i\coth\pi y=-i
     \left(1+\frac2{e^{2\pi y}-1}\right)
  \end{eqnarray}
can be reduced to the sum of the following contributions of the
integral over the segment of the real axis, $-m\leq z\leq m$, the
residue of the integrand at $z=+i\beta$ and the thermal-type
integral
  \begin{eqnarray}
    &&\frac1{2i}\int\limits_{{\mathbf C}_m} dz\,
      \frac{\cot\pi z}{(\beta^2+z^2)\sqrt{\alpha^2-z^2}}
      =-\frac{\pi}{\beta\sqrt{\alpha^2+\beta^2}}+
      \int_{-m}^{m}
      \frac{dx}{(\beta^2+x^2)
     \sqrt{\alpha^2-x^2}}\nonumber\\
   &&\qquad\qquad\qquad\qquad\qquad\qquad\qquad
      -2i\int\limits_0^\infty dy\,\frac1{e^{2\pi y}-1}\,
      \Big(f(y)-f(-y)\Big)~~,
  \end{eqnarray}
where the function $f(y)$ equals
 \begin{eqnarray}
   f(y)=\left.\frac1{(\beta^2+z^2)
     \sqrt{\alpha^2-z^2}}\,\right|_{\,z=m+iy}~~.  \label{2.18}
  \end{eqnarray}
Since
  \begin{eqnarray}
      \int_{-m}^{m}
      \frac{dx}{(\beta^2+x^2)
     \sqrt{\alpha^2-x^2}}=\frac2{\beta\sqrt{\alpha^2+\beta^2}}
   \arctan\left(\frac{\sqrt{\alpha^2+\beta^2}}\beta\,
        \frac{m}{\sqrt{\alpha^2-m^2}}\right)
  \end{eqnarray}
we finally have
  \begin{eqnarray}
    &&I(\alpha,\beta\,|\,m)=\frac\pi\beta\,
    \frac{\coth\pi\beta-1}{\sqrt{\alpha^2+\beta^2}}
      +\frac2{\beta\sqrt{\alpha^2+\beta^2}}
      \,\arctan\left(\frac{\sqrt{\alpha^2+\beta^2}}\beta\,
        \frac{m}{\sqrt{\alpha^2-m^2}}\right)\nonumber\\
      &&\qquad\qquad\qquad\qquad
      -2i\int\limits_0^\infty dy\,
      \frac1{e^{2\pi y}-1}\,
      \Big(f(y)-f(-y)\Big)~~.               \label{2.17}
  \end{eqnarray}
So far it was exact calculation without any approximation made. We are now
interested in the limit of large $\alpha$.
The calculation of the last integral in (\ref{2.17}) for $\alpha\gg1$
follows from the asymptotic expression for the combination of
functions $f(\pm y)$ in  (\ref{2.17})
 \begin{eqnarray}
   f(y)-f(-y)=\frac{i}{\sqrt{\alpha\delta}}\,
   \frac1{\alpha^2+\beta^2}\,
   \sqrt{\frac{(1+y^2/\delta^2)^{1/2}-1}
     {1+y^2/\delta^2}}\,
   \left(\,1+{\rm O}\left(\frac1{\alpha^{1/2}}
     \right)\right)~~,                           \label{2.20}
  \end{eqnarray}
where the parameter $\delta=\alpha-m$ is of order unity (bearing
in mind that $m$ will be identified with
$[\alpha]$). Thus the integral in (\ref{2.17}) after rescaling the
integration variable $y\to y\delta$ becomes
 \begin{eqnarray}
    -2i\int\limits_0^\infty dy\,
      \frac1{e^{2\pi y}-1}\,
      \Big(f(y)-f(-y)\Big)
      =\frac2{\alpha^2+\beta^2}
      \sqrt{\frac{\delta}\alpha}\,I(\delta)\,
      \left(\,1+{\rm O}\left(\frac1{\alpha^{1/2}}
        \right)\right)~~,
   \end{eqnarray}
where $I(\delta)$ is defined by Eq.(\ref{2.22}).

In the expression for the retarded potential (\ref{2.15}) the
parameter $m$ should be taken coinciding with
$[\alpha]$.  Then for integer values of
$\alpha$ the parameter $\delta$ will be vanishing and
$I(\delta)$ will be divergent. To isolate these
divergences we disentangle from the sum (\ref{2.16}),
$I(\alpha,\beta\,|\,[\alpha])$, the potentially divergent
terms at $n=\pm[\alpha]$,
   \begin{eqnarray}
     I(\alpha,\beta\,|\,[\alpha])
     =I(\alpha,\beta\,|\,[\alpha]-1)
     +\frac2{(\beta^2+[\alpha]^2)
       \sqrt{\alpha^2-[\alpha]^2}}~~.    \label{2.23}
  \end{eqnarray}
The last term here becomes singular at integer values of $\alpha$, but
the first term is absolutely finite, because its corresponding value of
$\delta=\alpha-[\alpha]+1$ belongs to the range $1\leq\delta<2$. Therefore,
using in (\ref{2.23}) the equation (\ref{2.17}) with $m=[\alpha]-1$ and
expanding the arctangent
   \begin{eqnarray}
   &&\frac2{\beta\sqrt{\alpha^2+\beta^2}}
      \,\arctan\left(\frac{\sqrt{\alpha^2
            +\beta^2}}\beta\,
        \frac{[\alpha]-1}{\sqrt{\alpha^2
            -([\alpha]-1)^2}}\right)\nonumber\\
      &&
      =\frac\pi{\beta\sqrt{\alpha^2+\beta^2}}
    -\frac{2^{3/2}}{\alpha^2+\beta^2}\,
    \sqrt{\frac{\alpha-[\alpha]+1}\alpha}
    +{\rm O}\left(\frac1{\alpha^{7/2}}\right)
  \end{eqnarray}
one finally obtains the expression (\ref{2.24}) used in Sect.4.

\end{document}